\documentclass[12pt]{article}
\usepackage{cite}
\usepackage{amsmath}
\usepackage{amsfonts}
\usepackage{epsf}
\usepackage{epsfig}
\usepackage{axodraw}
\numberwithin{equation}{section}
\newcommand{\beq}{\begin{eqnarray}}
\newcommand{\eeq}{\end{eqnarray}}
\newcommand{\bea}{\begin{eqnarray}}
\newcommand{\eea}{\end{eqnarray}}
\def\ba{\begin{eqnarray}}
\def\ea{\end{eqnarray}}
\def\nl{\nonumber\\}
\def\dz{\lambda_Z}

\newcommand{\litwo}{\mbox{Li}_2}
\newcommand{\lsim}{\raisebox{-0.13cm}{~\shortstack{$<$ \\[-0.07cm] $\sim$}}~}

\def\nl{\nonumber\\}
\def\nn{\nonumber}
\def\dz{\lambda_Z}
\def\dis{\displaystyle}
\renewcommand{\b}{{\mathbb B}}
\renewcommand{\c}{{\mathbb C}}
\newcommand{\cbar}{\bar{\mathbb C}}

\begin{document}
\thispagestyle{empty}
 \begin{flushleft}
DESY 00--127
\\
UG--FT--122/00
\\
hep-ph/0010193
\\
%begin 13 June 2000, 
October 2000 
%\\
%{\small version of {\today}, 
%/afs/ifh.de/group/theorie/riemann/fcnc/Journal-Article/article.tex}
\end{flushleft}
%-----------------------------------
 
 \noindent
 \vspace*{0.50cm}
\begin{center}
 \vspace*{1.5cm} 
{
%\huge  
\LARGE\bf
%Lepton Flavour Violation in $Z$ Decays
Charged Lepton Flavour Violation from

Massive Neutrinos in $Z$ Decays
}
% \vspace*{2.cm} 
 \vspace*{1.cm} 
%-----------------------------------
\\
{\large 
%J.I. Illana~$^{\$\ 1,2}$ 
J.I. Illana$^{a,b,*}$ 
and 
T. Riemann$^{a,\dag}$ 
}
\vspace*{0.5cm}
 
\begin{normalsize}
{\it
{\large $^{\ a}$}
Deutsches Elektronen-Synchrotron DESY
\\ 
Platanenallee 6, D-15738 Zeuthen, Germany

\bigskip

{\large $^{\ b}$}
Departamento de F{\'\i}sica Te\'orica y del 
Cosmos, Universidad de Granada
\\
Fuentenueva s/n, E-18071 Granada, Spain
}
\end{normalsize}
\end{center}
 
 \vspace*{1.5cm} 
 \vfill 

\begin{abstract}
Present evidences for neutrino masses and lepton flavour mixings allow to 
predict, in the Standard Model with light neutrinos, branching rates for the 
decays $Z \to e\mu, \mu\tau, e\tau$ of less than $10^{-54}$, while present 
experimental exclusion limits from LEP~1 are of order $10^{-5}$.
The GigaZ option of the TESLA Linear Collider project will extend the 
sensitivity down to about $10^{-8}$.
We study in a systematic way some minimal extensions
of the Standard Model and show that GigaZ might well be sensitive to 
the rates predicted from these scenarios.
\end{abstract}  

%------
 \normalsize
 \vfill
 \vspace*{.5cm}
  
  \bigskip
  \vfill
  \footnoterule
  \noindent

 {\small \noindent
%$^\dag$
%$^*$~Work supported in part  by the European Union under contract 
%             HPRN-CT-2000-00149.
%\\
%$^{\$}$~{On leave from Departamento de F{\'\i}sica Te\'orica y del 
%  Cosmos, Universidad de Granada, Fuentenueva s/n, E-18071 Granada, Spain.
%\\
%\hspace*{2mm}
$^*$~E-mail: {jillana@ugr.es}
\\
$^{\dag}$~E-mail: {riemann@ifh.de} 
%}
}
\newpage
%\tableofcontents
%\newpage
%  end of title page
%=========================================================== 
\section{\label{sec-intro}Introduction}
\setcounter{equation}{0}
%===========================================================
Lepton flavour violation searches are as old as our knowledge about the 
existence of at least two different kinds of leptons: electron and muon.
A prominent example of a {\em lepton flavour violating} (LFV) process is:
\ba
\mu \to e  \gamma.
\label{meg} 
\ea
This reaction has not been observed so far, 
and the best experimental upper limit of its branching fraction is
% due to the MEGA Collaboration at LAMPF 
\cite{Brooks:1999pu}: 
\ba 
{\rm BR}(\mu \to e \gamma) 
=
\frac{\Gamma(\mu \to e \gamma) }
%{\Gamma(Z\to{\rm all})}.
{\Gamma(\mu \to  \nu_{\mu}  e  \bar{\nu}_e)}  <  1.2 \times 10^{-11}.
\label{brmue}
\ea

At the $Z$ factory LEP, searches for quite similar LFV processes, but this
time directed to the $Z$ boson, became possible: 
\ba
Z \to e\mu, \mu\tau, e\tau .
\label{eq1}
\ea
The corresponding branching ratios are:
\ba
\label{lep-0}
{\rm BR}(Z\to \ell_1^{\mp}\ell_2^{\pm}) 
=
\frac{\Gamma(Z\to \bar \ell_1 \ell_2 + \ell_1 \bar \ell_2)} {\Gamma_Z},
\ea
and the best direct limits (95\% c.l.) are 
%due to OPAL and DELPHI 
\cite{PDG:1998aa}: 
\ba
\label{lep-em}
{\rm BR}(Z\to e^{\mp}\mu^{\pm}) &<& 1.7 \times 10^{-6}~~ 
\mbox{\cite{Akers:1995gz}},
\\
\label{lep-et}
{\rm BR}(Z\to e^{\mp}\tau^{\pm}) &<& 9.8 \times 10^{-6} ~~
\mbox{\cite{Adriani:1993sy,Akers:1995gz}},
\\
\label{lep-tm}
{\rm BR}(Z\to \mu^{\mp}\tau^{\pm}) &<& 1.2 \times 10^{-5} ~~
\mbox{\cite{Akers:1995gz,Abreu:1996mj}}.
\ea

These (and many other) observational facts may be described with the
concept of {\em lepton flavour conservation} (LFC) in neutral current reactions.
In the Standard Model of electroweak interactions (SM)
\cite{Glashow:1961ez,Weinberg:1967,Salam:1968rm}, 
lepton flavour is exactly conserved.
However, the model may be extended in such a way that virtual, LFC breaking 
corrections can appear.   
One mechanism relies on the assumption of neutrinos with finite masses and
lepton mixing (from a non-diagonal mass matrix of the gauge 
symmetry eigenstates) \cite{Pontecorvo:1957cp,Maki:1962mu,Pontecorvo:1968fh},
leading to tiny rates for all the above processes caused by LFV one-loop effects.
Historically, the $\nu$SM ---the Standard Model, enlarged  with
massive, mixing neutrinos--- was the first theory allowing  such predictions 
thanks to its renormalizability
\cite{Veltman:1968ki,'tHooft:1971rn,'tHooft:1972fi}.
For the reaction (\ref{meg}) and similar low-energy reactions like 
$\mu \to e^-e^+e^-$ or
$\nu_1 \to \nu_2  \gamma$ the first studies were reported in 
\cite{Petcov:1977ff,Cheng:1977uq,Marciano:1977wx}, and for the LFV $Z$ 
decays (\ref{eq1}) in \cite{Riemann:1982rq,Mann:1984dvt}.\footnote{
Soon later, related calculations were performed in the context of
flavour non-diagonal quark production with a heavy virtual top quark exchange
\cite{Riemann:1982sx,Ganapathi:1983xy,Clements:1983mk}. 
}

%====================================================================
\setlength{\unitlength}{1pt}
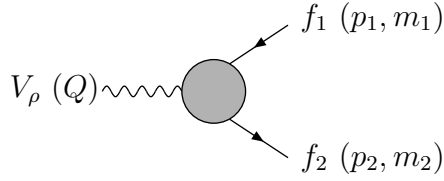
\begin{figure}[tbhp]
\begin{center}
\begin{picture}(100,60)(0,0)
\Photon(0,30)(30,30){2}{4.5}
\GOval(42,30)(12,12)(0){0.7}
\ArrowLine(70,55)(48,40.39)
\ArrowLine(48,19.51)(70,5)
\put(75,55){${\bar f}_1\ (p_1,m_1)$}
\put(75,2){$f_2\ (p_2,m_2)$}
%%%%\put(10,40){$q$}
%%%%\put(-5,22){$\mu$}
%%%%\put(-15,28){$V$}
\put(-35,28){$V_{\rho}\ (Q)$}
\end{picture}
\caption{\label{fig-effective}\it 
The effective LFV vertex.
}
\end{center}
\end{figure}
%====================================================================

The most general matrix element for the interaction of an on-shell 
vector boson with a fermionic current, as shown 
in Figure~\ref{fig-effective}, may be described by four dimensionless 
form factors.\footnote{
For an off-shell vector boson 
%the additional terms $(if_S+f_P\gamma_5)q^\nu$ may also contribute.
two more form factors contribute.
}
At one-loop order, it is convenient to parameterize 
\beq
{\cal M} =
-\frac{ig\alpha_W}{4\pi} \varepsilon^\rho \bar {\rm u}_{f_2}(p_2) 
%
%  \left[\gamma_\mu(f_V-f_A\gamma_5)
%     +\frac{1}{M_W}(if_M+f_E)\sigma_{\mu\nu}q^\nu
%      \right] 
%
\Gamma_{\rho}
{\rm u}_{f_1}(-p_1),
\label{M1}
\eeq
with $\alpha_W=g^2/(4\pi)$, $\varepsilon$ being the boson polarization vector 
and
\beq
\Gamma_{\rho}&=&
\gamma_\rho(f_V-f_A\gamma_5)
+ \frac{q^\nu}{M_W}(if_M+f_E\gamma_5)\sigma_{\rho\nu}
.
\label{M2}
\eeq
Above, $f_V$ and $f_A$ stand for vector and axial-vector couplings and 
$f_M$ and $f_E$ for magnetic and electric dipole moments/transitions of
equal/unlike final fermions.
The form factors depend on the  momentum transfer squared 
$Q^2=(p_2-p_1)^2$. 
For an on-shell photon, current conservation implies two additional 
conditions: $(m_2 - m_1) f_V = 0$ and $(m_2 + m_1) f_A = 0$.
This
means that LFV $\mu$ decays are exclusively due to dipole transitions, 
while for LFV $Z$ decays all $f_V, f_A, f_M, f_E$ are,
in principle, non-zero. 

The general expressions for the branching ratios are:\footnote{
For the quark flavour-changing $Z\to \bar q_1 q_2$, multiply by a colour factor 
$N_c=3$.}
\ba
\label{br1}
{\rm BR}(\mu \to e \gamma)  &=& \frac{12\alpha_W}{\pi} \frac{M^2_W}{m^2_\mu}
~\left(|f_M^\gamma|^2+|f_E^\gamma|^2\right),
%|{\cal V}_{\mu}(0)|^2
%\approx 2 \times 10^{-4} ~ \left(|f_M|^2+|f_E|^2\right)
%|{\cal V}_{\mu}(0)|^2, 
%\\
%\label{vbr1}
%|{\cal V}_{\mu}(0)|^2 &=& |f_M|^2+|f_E|^2,
\\
\label{br2}
{\rm BR}(Z\to \ell_1^{\mp}\ell_2^{\pm})  &=&
\frac{\alpha_W^3}{24\pi^2}
%========= \frac{\alpha^3}{192\pi^2 s_W^6c^2_W}
\frac{M_Z}{\Gamma_Z}
\left[|f_V^Z|^2+|f_A^Z|^2+
\frac{1}{2c^2_W}\left(|f_M^Z|^2+|f_E^Z|^2\right)\right].
%~|{\cal V}(M^2_Z)|^2
%\approx 10^{-6}~
%\left[f_V|^2+|f_A|^2+\frac{1}{2c^2_W}\left(|f_M|^2+|f_E|^2\right)\right].
%|{\cal V}(M^2_Z)|^2
\label{vbr2}
%|{\cal V}(M^2_Z)|^2 &=& 
%|f_V|^2+|f_A|^2+\frac{1}{2c^2_W}\left(|f_M|^2+|f_E|^2\right).
\ea
%Further, as will be shown in Section \ref{sec-SM}, it is:
%The sums of squared form factors depend on the squared momentum transfer $Q^2$,
Notice that while the muon total width is 
%$\propto \alpha^2 m^5_\mu/M^4_W$, the $Z$ has $\Gamma_Z \propto \alpha M_Z$, 
$\Gamma_\mu=\alpha^2_W/(384\pi) \ m^5_\mu/M^4_W$, the $Z$ width is
$\Gamma_Z \approx \alpha_W/c^2_W \ M_Z$.
That is why ${\rm BR}(Z\to\ell_1^{\mp}\ell_2^{\pm})$ is naturally by an 
order of $\alpha_W$ smaller than ${\rm BR}(\mu \to e \gamma)$. 
Furthermore, the $M^2_W/m^2_\mu$ enhancement of (\ref{br1})
is compensated due to the chirality-flipping character of the dipole form
factors, proportional to the fermion mass $m_\mu$.

The form factors are model-dependent. 
In the approximation of {\em massless electrons} (for $\mu\to e\gamma$) or 
{\em massless external leptons} (for $Z\to\bar\ell_1\ell_2$), there is 
{\em only one independent form factor} in each case. In the simplest assumption 
of $n$ Dirac virtual neutrinos $\nu_i$ with masses $m_i$, the mixings factor out
and one can write
\ba
\mu\to e\gamma:&&
f_M^\gamma = f_E^\gamma \equiv \frac{s_W}{16}\frac{m_\mu}{M_W}
\sum_{i=1}^n {\bf V}_{\mu i} {\bf V}_{e i}^*
V_\gamma(\lambda_i;\lambda_Q),\\
Z\to \bar\ell_1 \ell_2:&&
f_V^Z = f_A^Z \equiv 
\frac{1}{4c_W}\sum_{i=1}^n {\bf V}_{\ell_1 i} {\bf V}_{\ell_2 i}^*
V_Z(\lambda_i;\lambda_Q),
\label{order-V}
\\
&& f_M^Z=f_E^Z=0,
\ea
where ${\bf V}$ is the lepton-flavour mixing matrix and $V_{\gamma/Z}$ are
vertex functions, fully describing the amplitudes. We have introduced the 
neutrino mass ratios $\lambda_i = m_i^2/M_W^2$ and the virtuality of the $Z$ 
boson $\lambda_Q = Q^2/M^2_W$, that becomes $\lambda_Z = M_Z^2/M_W^2$ on its 
mass shell.\footnote{
The values
$M_W = 80.41 ~ {\rm GeV}$,
$M_Z = 91.187  ~ {\rm GeV}$,
$c_W=M_W/M_Z$, $g=e/s_W$,  and
$\Gamma_{Z} =  2.49 ~ {\rm GeV}$ will be taken throughout this work.}
Owing to the unitarity of the mixing matrix, the amplitudes vanish for massless 
or degenerate virtual neutrinos, in exact analogy with the GIM cancellation 
in the quark sector \cite{Glashow:1970st}. 
%=================================================
%\bigskip

%Let us now have a closer look at the differences between LFV $\mu$ and $Z$ 
%decays. 
We have strong evidence for neutrino masses of the order of some 
fractions of eV and large mixings \cite{Mann:1999zb,Bahcall:1998jt}.
For small neutrino masses, a power-series expansion of the muon decay 
amplitude yields \cite{Petcov:1977ff,Cheng:1977uq,Marciano:1977wx}:
\ba
\label{apprmu}
V_\gamma(\lambda_i \ll 1; 0) \approx 
%\frac{10}{3} -
\mbox{const}_\gamma+
\lambda_i + {\cal O}(\lambda_i^2),
\ea
and similarly for the $Z$ decay,\footnote{
This is in clear 
distinction to Eqn. (6) of \cite{Pham:1998xhp} (with a logarithmic mass 
dependence), where from the recent neutrino data a prediction was derived  
to be
${\rm BR}(Z\to\mu^{\mp}\tau^{\pm} )\approx {\cal O}(10^{-8} \div
10^{-5})$.
} 
but with complex coefficients
\cite{Riemann:1982rq,Illana:1999ww,Illana:2000zy}: 
\ba
\label{apprZ}
% V_Z(m_i^2/M_W^2 \ll 1; M_Z^2/M_W^2=1.25) \approx (1.25 + 1.03~ i) + (2.53 -
% 2.31 ~ i) \frac{m_i^2}{M_W^2} + \ldots.
V_Z(\lambda_i \ll 1; \lambda_Z) \approx  \mbox{const}_Z + (2.562 -
2.295 ~ i) \lambda_i + 
{\cal O}(\lambda_i^2).
\ea
The constant terms drop out after summing over the $n$ generations of
mixing neutrinos, but there survive contributions to the branching fractions 
proportional to the fourth power of the mass ratio $m_i/M_W$, for
non-degenerate neutrinos, and thus unfortunately very small.
%See also GIM \cite {Glashow:1970st}.
%Of course, due to this property, 
Therefore, an observation of such LFV decays would be 
indicative to the existence of New Physics with a new, large  mass scale 
involved. 

Consider now the hypothetical case of large neutrino masses.
Neutrinos with large masses are accommodated by many extensions of the SM like 
grand unified theories \cite{Langacker:1981js} or superstring-inspired models 
with an $E_6$ symmetry \cite{Hewett:1989xc}. Heavy neutrinos are also well 
motivated by the seesaw mechanism 
\cite{Yanagida:1980xy,Gell-Mann:1980vs,Mohapatra:1980ia}.
{From} the exact expression of the LFV $\mu$ decays \cite{Langacker:1988up}: 
\begin{eqnarray}
  \label{eq:intlar}
% F_\gamma(m_i^2/M_W^2 \gg 1; Q^2/M^2_W) &\approx&  \frac{4}{3} + 8 
% \frac{\ln(m_i^2/M_W^2)}{m_i^2/M_W^2} 
 V_\gamma(\lambda_i ; 0) &=&  
 \mbox{Const}_\gamma
 + \Phi(\lambda_i);
\\
\Phi(x)&=&\frac{x(1-6x+3x^2+2x^3-6x^2\ln x)}{(1-x)^4},
\end{eqnarray}
one obtains $\Phi(x\gg 1)\to 2$. In contrast, for the LFV $Z$ decays
\cite{Mann:1984dvt}:
\begin{eqnarray}
  \label{eq:intla}
  V_Z(\lambda_i \gg 1; \lambda_Q) &\approx& \mbox{Const}_Z +
  \frac{\lambda_i}{2} + 
  {\cal O}(\ln\lambda_i).
\end{eqnarray}

%It is known for a long term that the see-saw mechanism 
%\cite{Yanagida:1980xy,Gell-Mann:1980vs,Mohapatra:1980ia}
%with a general Majorana 
%\cite{Majorana:1937vz} mixing 
%matrix may be a natural origin of both small {\em and} large masses,
%but at the price of 
%small mixings if there are large mass splittings.
%However, this may be circumvented with inter-generation mixings
%\cite{Bernabeu:1993up,Korner:1993an,
%}, see also: \cite{
%Ilakovac:1995kj}.

Let us summarize the phenomenologically relevant differences 
between the LFV $\mu$ and $Z$ decays:
(i) the very different origin of the form factors intervening (dipoles in
    the $\mu$ case 
    %(\ref{br1}) 
    and mostly vector and axial-vector
    in the $Z$ case);
    %(\ref{vbr2});
(ii) the `typical size' of the rates due to the different powers of the 
     coupling constant $\alpha_W$ appearing in the branching fractions;
and 
(iii) for fixed mixings, the $Z$ branching ratio rises with virtual 
      neutrino masses while the $\mu$ branching ratio reaches a plateau.
%     (compare Eqns. (\ref{eq:intlar}) and (\ref{eq:intla})).

%\bigskip

In the rest of this work, we will concentrate on one-loop induced
LFV $Z$ decays. For these and other rare $Z$ decays, the branching fractions 
are typically
\ba
\label{2loop}
 {\rm BR}(Z\to \mbox{rare}) \sim \left( \frac{\alpha_W}{\pi}
 \right)^2 \sim 
 {\cal O}(10^{-6}) .
\ea
There are many 
%LEP~1 
studies on such processes, in relation to e.g. CP violation
\cite{Bernabeu:1986pk,Rius:1990gk},
heavy neutral singlets 
\cite{Gonzalez-Garcia:1990fb,Dittmar:1990yg
%,Dittmar:1990mn
},
supersymmetry 
\cite{Duncan:1985vt,Gabbiani:1988pp}
and superstrings
\cite{Bernabeu:1987gr,delAguila:1988nn} or
%LFV $Z$ decays 
induced by a mixing with a heavy $Z'$ 
%are studied in 
\cite{Langacker:2000ju}.
See also the summary report of the LEP~1 Workshop \cite{Bernreuther:1989rt}
and the later study on the high luminosity LEP~1 project
\cite{Blucher:1991hm}, in particular
\cite{Dittmar:1990mn}.
% in \cite{Bergstrom:1991ra}.
The discovery reach of LEP~1 was indeed not very large, after comparing the
experimental limits (\ref{lep-em})--(\ref{lep-tm}) with the order of 
magnitude of the potential effects (\ref{2loop}). 

In a few years from now, a new high energy $e^+e^-$ Linear Collider could be 
constructed. 
Interesting enough, with the GigaZ option of the TESLA Linear Collider project 
one may expect the production of about $10^{9}$ $Z$ bosons at resonance 
\cite{Hawkings:1999ac}.
This huge rate, about a factor 1000 higher than the one at LEP~1, will
make possible checks of the SM and its minimal supersymmetric extension MSSM 
at the two-loop
level \cite{Erler:2000jg}, as well as searches for any kind of rare $Z$ decays
with unprecedented precision. A careful analysis \cite{Wilson:1998bb} shows 
that in particular the LEP~1 discovery limits could be reduced to
\ba
\label{lep-emx}
{\rm BR}(Z\to e^{\mp}\mu^{\pm}) &<&  2 \times 10^{-9},
\\
\label{lep-etx}
{\rm BR}(Z\to e^{\mp}\tau^{\pm}) &<& \kappa \times 6.5 \times 10^{-8},
\\
\label{lep-tmx}
{\rm BR}(Z\to \mu^{\mp}\tau^{\pm}) &<& \kappa \times 2.2 \times 10^{-8},
\ea
with $\kappa = 0.2 \div 1.0$.  
This means one might have a chance of observation if the lepton mixings 
are not tiny and the masses of the neutrinos are at least of the order of the 
weak scale. Furthermore, in view of the expected sensitivities it 
might well be that the predictions are such that not only the asymptotic limit 
for large internal masses but {\em an exact calculation of the effective 
vertex is needed}: at least, it will be important to know where 
the large-mass limit fails. 

We perform a complete recalculation of the branching ratio 
(\ref{lep-0}) in presence of heavy Dirac or Majorana neutrinos and 
study the prospects for GigaZ in view of present, related experimental facts.
We also compare to earlier studies and revise some of them.
Many technical details of more pedagogical character may be found in
\cite{Illana:1999ww}.
In Section 2 the case of Dirac neutrinos is considered; Majorana neutrinos
are treated in Section 3 and our conclusions are drawn in Section 4.
The Appendix collects notations, conventions and useful expressions for
the tensor integrals and the vertex functions as well as their low and 
large neutrino-mass limits.

%=====================================================================
\section{\label{s-lfv}The LFV $Z$ decays in the  $\nu$SM
}
%=====================================================================
The simplest extension of the SM accounting for non-vanishing LFV $Z$ decay
rates consists of extending the particle content of the SM with three 
right-handed $\nu$ singlets, thus forming three massive, mixing neutrino states 
\`a la Kobayashi-Maskawa. This is in conformity with compatible results from 
present solar, atmospheric, reactor and accelerator neutrino experiments.

On basically the same footing one may also study the case of an 
additional sequential, but heavy neutrino state.
This case implies the existence of a heavy charged lepton as well,
in order to keep total lepton number $L$ conserved.\footnote{
A fourth generation of quarks is also needed to keep the
theory anomaly free.} 
It is not a very
favoured scenario but we consider it as a simple application. 

\begin{figure}[tbhp]
D1:\put(10,-30){\epsfig{file=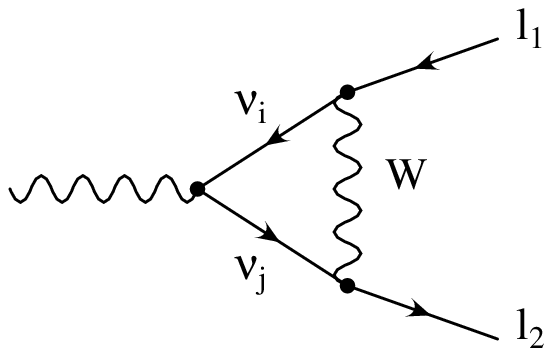,angle=0,width=0.25\linewidth}}
\hspace{4.5cm}
D2:\put(10,-30){\epsfig{file=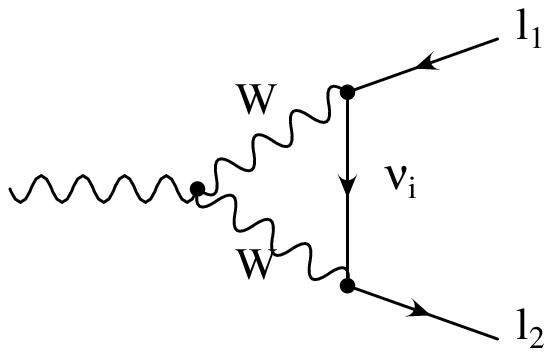,angle=0,width=0.25\linewidth}}
\hspace{4.5cm}
D3:\put(10,-30){\epsfig{file=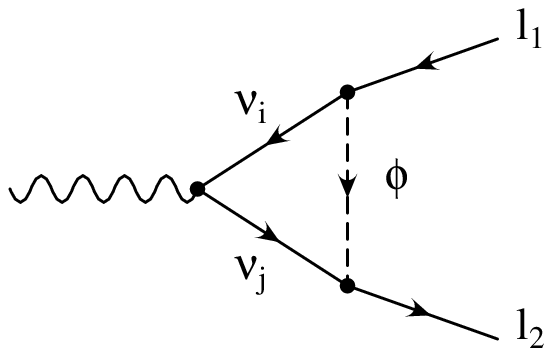,angle=0,width=0.25\linewidth}} \\
D4:\put(10,-30){\epsfig{file=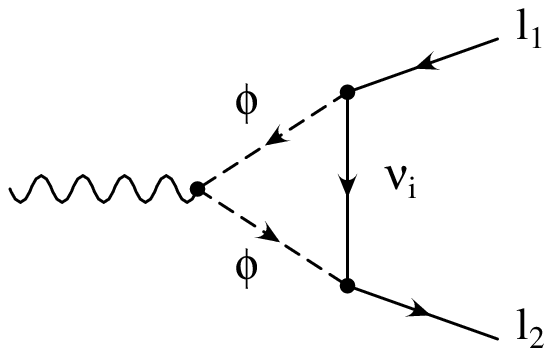,angle=0,width=0.25\linewidth}}
\hspace{4.5cm}
D5:\put(10,-30){\epsfig{file=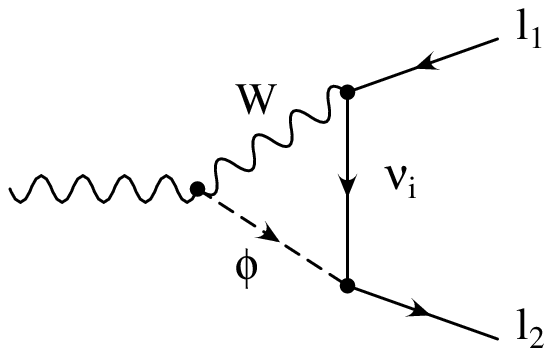,angle=0,width=0.25\linewidth}}
\hspace{4.3cm} + crossed \\
D$\Sigma$:
 \put(10,-23){\epsfig{file=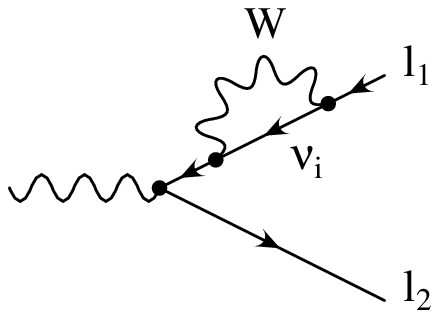,angle=0,width=0.2\linewidth}}
\hspace{3.3cm}
+\put(10,-23){\epsfig{file=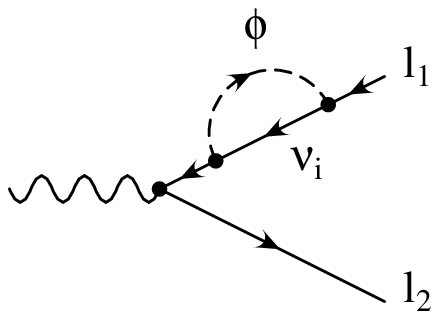,angle=0,width=0.2\linewidth}}
\hspace{3.3cm}
+\put(10,-36){\epsfig{file=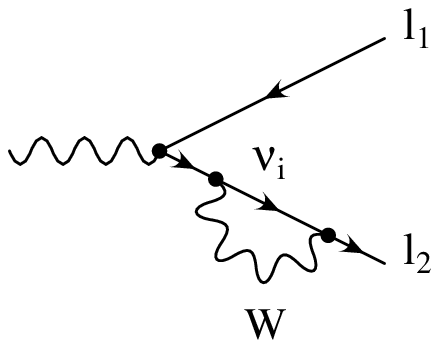,angle=0,width=0.2\linewidth}}
\hspace{3.3cm}
+\put(10,-38){\epsfig{file=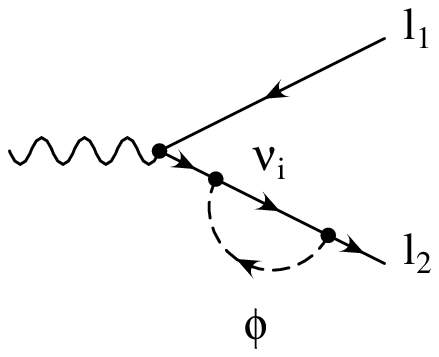,angle=0,width=0.2\linewidth}}
\caption{\label{fig-vertex}   \it 
Feynman diagrams for the lepton-flavour changing $Z$ decay. 
In the case of virtual, ordinary Dirac neutrinos, 
the $Z\nu_i\nu_j$ vertices in D1 and D3 
are flavour-diagonal.
The analogous quark-flavour changing process can be
obtained by replacing $l_k$ by down-quarks and $\nu_i$ by up-quarks. 
}
\end{figure}

The final state charged leptons may be assumed massless.
The amplitude is then purely left-handed and it is described by a single form 
factor,
\beq
{\cal M}&=&-\frac{ig\alpha_W}{16\pi c_W}
{\cal V}(Q^2)\
\varepsilon^\rho_Z \bar {\rm u}_{\ell_2}(p_2)\gamma_\rho(1-\gamma_5) 
{\rm u}_{\ell_1}(-p_1)
.
\label{ampli}
\eeq
%For Dirac neutrinos the mixings factor out. 
Using the same vertex function $V_Z$ introduced in (\ref{order-V}) one has:
\beq
{\cal V}_{\rm Dir}(Q^2)&=&
\sum_{i=1}^n {\bf V}_{\ell_1i} {\bf V}_{\ell_2i}^* V_Z(\lambda_i;\lambda_Q)
.
\eeq
In the 't Hooft Feynman gauge, the amplitude receives contributions
from the set of diagrams of Figure~\ref{fig-vertex}:
\ba
\label{l24}
V_Z(\lambda_i;\lambda_Q) &=&
%\left[ 
v_W(i)+v_{WW}(i)+v_{\phi}(i)+v_{\phi\phi}(i)+v_{W\phi}(i)+v_{\Sigma}(i)
%\right]
.
\ea
The vertex diagrams D1 to D5 yield respectively:
\beq
\label{D1}
v_W(i) & = & 
-(v_i+a_i)\ \left[ \lambda_Q\ (\c_{0}+\c_{11}+\c_{12}+\c_{23})-2\c_{24}+1\right] 
\nonumber\\&   & 
-(v_i-a_i)\ \lambda_i\   \c_{0},
\\ 
\label{D2}
v_{WW}(i)  & = & 2c^2_W\ (2I^{i_L}_3) \left[ \lambda_Q\ 
({\cbar}_{11}+{\cbar}_{12}+{\cbar}_{23})-6{\cbar}_{24}+1\right],
\\  
\label{D3}
 v_{\phi}(i)  & = &-(v_i+a_i)\ 
\displaystyle\frac{\lambda^2_i}{2}\  \c_0 
%\nonumber\\ && 
-(v_i-a_i)\ \displaystyle\frac{\lambda_i}{2}\ 
          \left[\lambda_Q\ \c_{23} - 2 \c_{24} + \displaystyle\frac{1}{2} \right],
\\ 
 \label{D4}
v_{\phi\phi}(i)  & = &-(1-2s^2_W)\ (2I^{i_L}_3)\ \lambda_i\
{\cbar}_{24},
\\  
\label{D5}
v_{W\phi}(i)  & = & -2s^2_W\ (2I^{i_L}_3)\ \lambda_i\ {\cbar}_0.
\eeq

%=====================================================================

\noindent
The self-energy corrections to the external fermion lines D$\Sigma$ contribute
with:
\beq
\label{Dse}
v_{\Sigma}(i)  & = & 
\displaystyle\frac{1}{2}(v_i+a_i-4c^2_W a_i)\left[(2+\lambda_i)\b_1 +1 \right].
\eeq
The definitions of weak neutral vector and axial-vector couplings are as usual:
\ba
\label{def-v}
v_i &=& I^{i_L}_3 - 2 Q_i s_W^2 = I^{i_L}_3(1-4 s^2_W |Q_i|),
\\
\label{def-a}
a_i &=& I^{i_L}_3,
\ea
and the dimensionless one-loop tensor integrals 
$\c_0$, $\cbar_0$, $\c_{ab}$, $\cbar_{ab}$ and $\b_1$ 
are given in Appendix \ref{a-vert}, taking arguments $\lambda_i=\lambda_j$
for the $\c$ functions.

The form factor ${\cal V}$ describing the amplitude (\ref{ampli}) is finite and 
no renormalization is needed, as expected because there is no tree-level 
coupling of a $Z$ boson to two fermions of different flavour. Nonetheless,
a non-trivial cancellation of infinities takes place, since $\c_{24}$,
$\cbar_{24}$ and $\b_1$ are UV-divergent. Actually, the vertex function
$V_Z(\lambda_i;\lambda_Q)$ is still infinite but has {\em divergences 
independent of} $\lambda_i$, that makes possible the cancellation of
the divergent terms in the amplitude, thanks to unitarity of the mixing matrix.
 
%=====================================================================
\subsection{\label{s-light}Contributions from light neutrinos
}
%=====================================================================
Disregarding the controversial results of the LSND accelerator 
experiment, all neutrino experiments are compatible with the 
oscillation  of three neutrino species. 
We will now estimate the LFV branching ratios under the assumption that
there are three generations of light neutrino flavours and that their mixing is
given by the unitary mixing matrix ${\bf V}$ constrained by the experiments.
The mixing is described by three angles $\vartheta_{12}, \vartheta_{13},
\vartheta_{23}$, and one CP-violating phase $\delta$ as in the 
quark CKM case.\footnote{
Oscillation experiments cannot distinguish between the Dirac or Majorana
character of the neutrinos. If they happen to be Majorana particles,
two additional CP-violating `Majorana' phases $\alpha, \beta$ are needed
since for  {\em strictly} neutral particles less phase factors may be `eaten' 
by redefining complex fermion fields. They are set here to $\alpha=\beta=0$.}   

A global analysis of atmospheric neutrino data favours
$\nu_{\mu}-\nu_{\tau}$ oscillations \cite{Gonzalez-Garcia:2000sq},
\bea
\Delta m^2_{\rm atm} &=& \Delta m^2_{23} \simeq 
(1 \div 6) \times 10^{-3} {\rm eV}^2,
\label{atmos}
\\
\sin^2 2\vartheta_{\rm atm} &=& \sin^2 2\vartheta_{23} \simeq 
0.8 \div 1.0.
\eea
The solar neutrino deficit is compatible with $\nu_{e}-\nu_{\mu}$ oscillations
\cite{Gonzalez-Garcia:2000sq}, 
\bea
\Delta m^2_{\odot} &=& \Delta m^2_{12} \simeq 
10^{-10} \div 10^{-5} {\rm eV}^2 ,
\label{solar}
\\
\sin^2 2\vartheta_{\odot} &=& \sin^2 2\vartheta_{12}=\mbox{free} .
\eea
There are solutions for vacuum and matter oscillations compatible
with a wide range of masses and mixing angles, although the large mixing 
angle solution LMA with maximal mass splitting seems favoured. 
From reactor searches, there are no 
hints of $\nu_{e}-\nu_{\tau}$ oscillations \cite{Apollonio:1999ae}, which 
allows us to assume
\bea
\sin^2 2\vartheta_{13} = 0.
\label{chooz}
\eea
Taking this information into the standard parameterization for the mixing
matrix \cite{PDG:1998aa} 
%and putting the Dirac CP-phase $\delta=0$, as no information on it is yet 
%available, 
one has
\bea
{\bf V} =
\label{mixmat}
\begin{pmatrix}
c_{12} & s_{12} & 0
\\
-\frac{1}{\sqrt{2}}s_{12} & \frac{1}{\sqrt{2}}c_{12} & \frac{1}{\sqrt{2}}
\\
\frac{1}{\sqrt{2}}s_{12} & -\frac{1}{\sqrt{2}}c_{12}&\frac{1}{\sqrt{2}}
\end{pmatrix}.
\eea

Using the unitarity of ${\bf V}$  and $\ell_1\neq\ell_2$,
\ba
{\rm BR}(Z\to \ell_1^{\mp}\ell_2^{\pm})
=\frac{\alpha_W^3M_Z}{192\pi^2c_W^2\Gamma_Z}
%\nn\\
\ \left|\sum^3_{i=1} {\bf V}_{\ell_1 i} {\bf V}^*_{\ell_2 i}
[V_Z(\lambda_i,\lambda_Z)-V_Z(0,\lambda_Z)]\right|^2.
\ea
Performing a well justified low neutrino mass expansion of the vertex function 
(see Appendix \ref{a-zeromass}), one finds \cite{Riemann:1982rq,Illana:1999ww}:
\ba
\label{vlami}
V_Z(\lambda_i,\lambda_Z)-V_Z(0,\lambda_Z) &=& a_1 \lambda_i 
+ {\cal O}(\lambda^2_i),
\\
\label{a1reim}
a_1 &=& 2.5623 - 2.2950\ i.
\ea
Therefore BR$(Z\to \ell_1^{\mp}\ell_2^{\pm})$ goes as $m^4_i$
for low neutrino masses.  
%
%=========================================================================
\begin{figure}[th]
\begin{center}
\epsfig{file=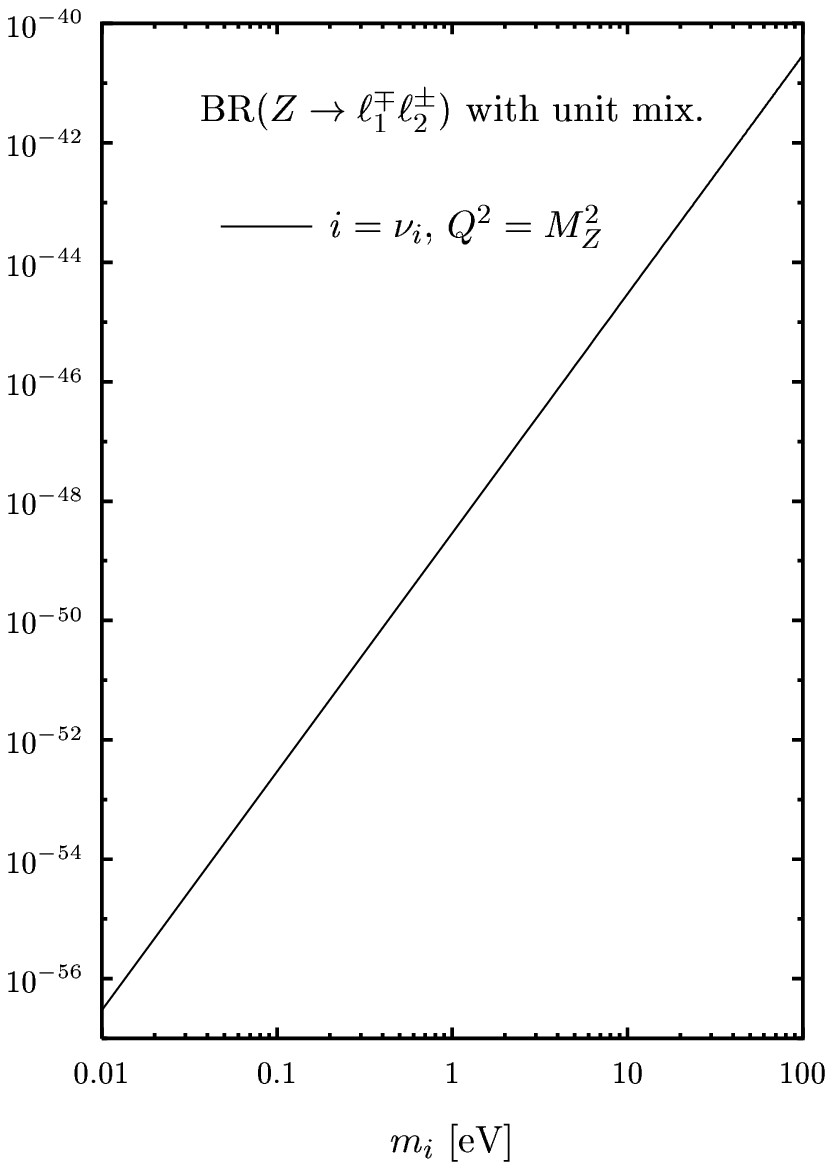,angle=0,width=0.49\linewidth}
\epsfig{file=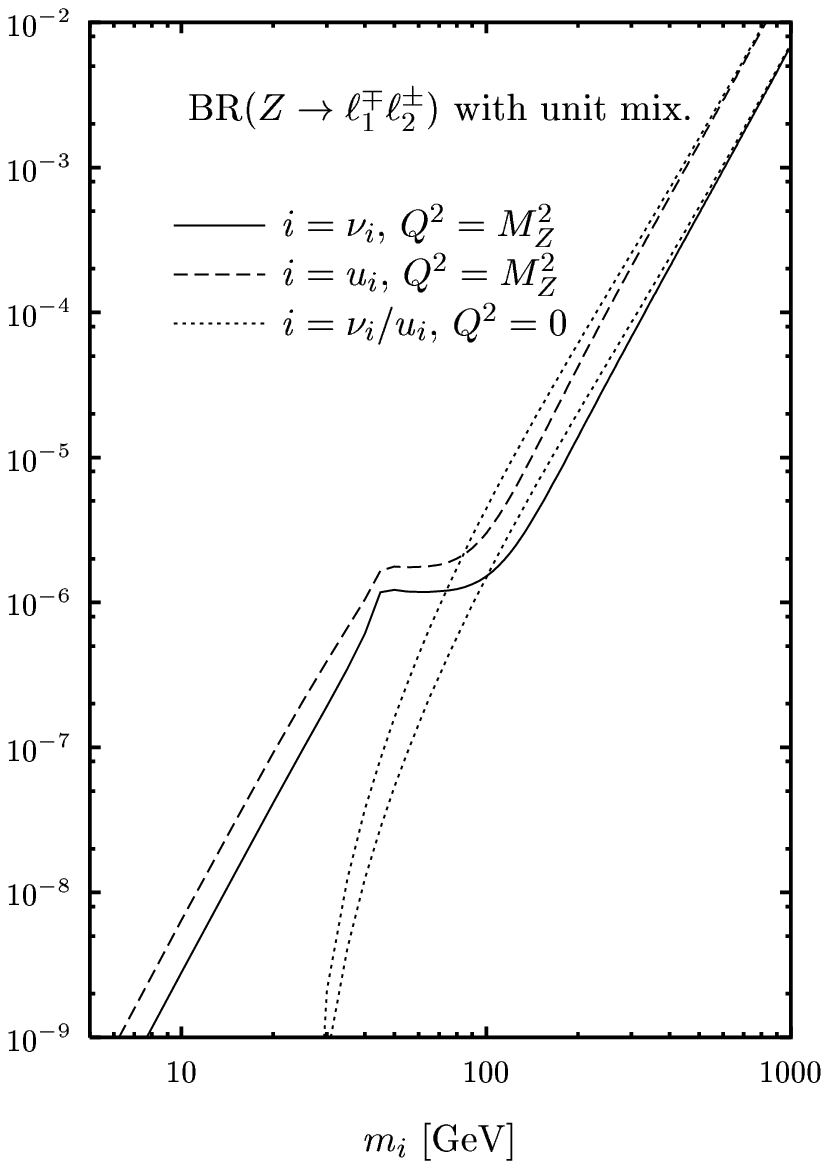,angle=0,width=0.49\linewidth}
\end{center}
\caption{
\label{figure-dirac}
{\it 
Contribution of one neutrino generation $i$ to the LFV 
$Z\to\ell_1^\mp\ell_2^\pm$
decays for ordinary Dirac neutrinos in the small and large neutrino mass regions,
and the analogous quark case. The mixing factor has been set to
${\bf V}_{l_1 i}{\bf V}^{*}_{l_2 i}=1$.}
}
\end{figure}
%=========================================================================
%
This behaviour is shown in Figure \ref{figure-dirac}.
It is valid over a large mass range until about $m_i \approx 30$ GeV, 
i.e. just below the $Z$ mass. 

Taking now into account the phenomenological squared mass differences 
$\lambda_{ij}\equiv\Delta m^2_{ij}/M^2_W$ and the mixing angles
(\ref{atmos})--(\ref{chooz}), one can determine the finite expectation:
\ba
\label{predfin}
{\rm BR}(Z\to \mu^{\mp}\tau^{\pm}) \simeq
3\times 10^{-6}\times
|s^2_{12}\lambda_{12}-\lambda_{23}|^2 
\approx 
(1\div 30)\times 10^{-55},
\ea
and the upper limit:
\ba
\label{predlim}
{\rm BR}(Z\to e^{\mp}\mu^{\pm}) 
\simeq
{\rm BR}(Z\to e^{\mp}\tau^{\pm})
\approx
6\times 10^{-6}\times c^2_{12}s^2_{12}\lambda^2_{12} 
\lsim
4\times 10^{-60}.
\ea 
These extremely small rates are far beyond experimental verification.
This justifies taking the light neutrino sector as massless in the following 
sections where we discuss extensions providing larger rates.

%=====================================================================
\subsection{\label{s-dirac}Contributions from one heavy ordinary Dirac 
neutrino 
}
%=====================================================================
Assume the neutrino of generation $N$ to be the only heavy one, mixing with
a light sector with negligible masses. Then, using again the unitarity of the 
mixing matrix:
\bea
{\rm BR}(Z\to \ell_1^{\mp}\ell_2^{\pm})&=& 
\frac{\alpha_W^3M_Z}{192\pi^2c_W^2\Gamma_Z}
\left| {\bf V}_{\ell_1N}{\bf V}_{\ell_2N}^*\right|^2
%\times
\left|V_Z(\lambda_N;\lambda_Z) - V_Z(0;\lambda_Z)\right|^2.
\label{2.23}
\eea
In the large Dirac neutrino mass limit, the following approximation works well
(see Appendix \ref{appsub-maj}):
\ba
\label{vbig}
V_Z(\lambda;\lambda_Z)
&=&
\frac{1}{2}
\Biggl[
%\frac{8c_W^2}{D-4}
-4c^2_W\Delta_\epsilon
+
\lambda
+
\left(3-\frac{\lambda_Z}{6} (1-2s_W^2) \right) \ln\lambda
\nl
&&+~\frac{1}{18} \left(-66 - \lambda_Z + 96 s_W^2 +5 s_W^2\lambda_Z \right)
\\
%\nl 
&&+~\frac{1}{3}
% in Ann.Phys. we had 32/8s_W^2\dzy\arctan\left(\frac{1}{2y}\right) instead of 32/3.
% this was wrong, but only a typo. see notes of 1982 in VQ82, 25-11-82-7.
\left(-8+2\lambda_Z-32 s_W^2-4s_W^2\lambda_Z\right) y 
\arctan\left(\frac{1}{2y}\right)
%%%%%%%%%% y\arctan\left(\frac{1}{2y}\right) 
%\nl 
%&&
%%-~ \frac{7}{36}  s_W^2 |Q_i|\lambda_Q             
\Biggr]
+ {\cal O}\left(\frac{\ln\lambda}{\lambda}\right),
\nonumber
\ea
with 
\ba
y &=& \sqrt{1/\dz  - 1/4}.
\label{y}
\ea
The vertex function contains a constant term proportional to 
$\Delta_\epsilon=2/(4-d)+\gamma+\ln4\pi$, divergent in $d=4$ dimensions.
This term drops out in the physical amplitude, as expected, since the 
unitarity of the mixing matrix demands the subtraction of $V_Z(0;\lambda_Q)$,
with identical divergences.
Its expression can be found in Appendix \ref{a-zeromass}.
For an on-shell $Z$,
\ba
\label{bigva0}
V_Z(\lambda_N;\lambda_Z) - V_Z(0;\lambda_Z)
=
\frac{1}{2}
\left[
\lambda_N + 2.88 \ln \lambda_N 
%- 0.05 |Q_{\nu}|
%\right]
%\nl &&
-
(6.99 + 2.11 \ i) 
% - (0.23 + 0.41 \ i) |Q_{\nu}|   
\right] +{\cal O}(\ln\lambda_N/\lambda_N).\hspace{4mm}
\label{d40}
\ea
The exact results are depicted in Figure \ref{figure-dirac}, where the
simpler calculation with $Q^2=0$ \cite{Ma:1980px} is also displayed.
We find agreement with earlier calculations \cite{Mann:1984dvt}, 
also for quark flavour-changing $Z$ decays 
\cite{Ganapathi:1983xy,Clements:1983mk}.
%The quantity presented is related to a branching ratio definition
%often used in the literature:
%%
%\ba
%\label{bsubz}
%B_Z &\equiv& \frac{\Gamma(Z\to f_1^{\mp}f_2^{\pm}
%%\bar l_1 l_2 + l_1 \bar l_2
%)}
%                          {2 \times \Gamma(Z\to \bar \nu_l \nu_l)}
%= \left(\frac{\alpha}{\pi}\right)^2\frac{N_c}{16s^4_W}
%~|{\cal V}(M^2_Z)|^2, 
%\ea
%with $N_c$ as colour factor.
%At very large mass, $m_N^2 \gg M_Z^2$, the approximation of $V$ with 
%$ M_Z^2 = 0$ works well, while it naturally breaks down completely for 
%smaller neutrino masses.
Of course, the results for $Q^2=0$ are a good approximation only when 
$m_N^2 \gg M_Z^2$.

For a study of the size of the branching ratios, the knowledge of the 
light-heavy mixing elements involved in (\ref{2.23}) is crucial. 
Their values do not only influence potential LFV processes but also 
flavour-diagonal ones.
Using a general formalism developed in \cite{Langacker:1988ur} one can
exploit measurements of flavour diagonal processes (checks of lepton 
universality and CKM unitarity, $Z$ boson invisible width, etc.) 
\cite{Nardi:1992rg,Nardi:1994iv} to obtain {\em indirect} experimental 
bounds on such light-heavy mixings \cite{Ilakovac:1995kj}, defined as
\bea
s^2_{\nu_\ell} \equiv |\sum_{i} {\bf V}_{\ell N_i}|^2.
\label{lhdef}
\eea
The most recent indirect bounds \cite{Bergmann:1998rg}:
\bea
\label{lhmixne}
s^2_{\nu_e} &<&   0.012,
\\ 
\label{lhmixnm}
s^2_{\nu_\mu}&<&0.0096,
\\ 
s^2_{\nu_\tau}&<&0.016,
\label{lhmixnt}
\eea
are only improved by the impressive accuracy of the {\em direct} searches for 
LFV processes involving the first two lepton generations. 
In fact, for heavy enough neutrinos one can rewrite (\ref{br1}):
\ba
{\rm BR}(\mu\to e\gamma)\approx\frac{3\alpha}{8\pi} s^2_{\nu_e}s^2_{\nu_\mu},
\label{plateau}
\ea
and from (\ref{brmue}) a stringent `mass-independent' limit can be extracted
\cite{Tommasini:1995ii}:
\bea
\label{lhmix}
s^2_{\nu_e}s^2_{\nu_\mu}<1.4\times 10^{-8}.
\eea

The limit above sends the $Z\to e\mu$ process beyond any experimental reach, 
even if the neutrinos are very heavy. 
At this point, it is important to realize that, 
although the branching fractions for large neutrino masses grow as 
\bea
{\rm BR}(Z\to \ell_1 \ell_2) \propto s^2_{\nu_{\ell_1}} s^2_{\nu_{\ell_2}} 
m^4_N,
\eea
there is a `natural' upper value for the neutrino mass determined by
the perturbative unitarity condition on the heavy Dirac neutrino decay width
\cite{Chanowitz:1979mv,Durand:1990zs,Durand:1992wb,Fajfer:1998px,%
Ilakovac:1999md},
\beq
\Gamma_{N_i}\simeq
\frac{\alpha_W}{8M^2_W}m^3_{N_i}\sum_{\ell}|{\bf V}_{\ell N_i}|^2\leq 
\frac{1}{2} m_{N_i},
\label{pub}
\eeq
that implies
\beq
m^2_{N}&\leq&\frac{4 M^2_W}{\alpha_W}
\left[\sum^{3}_{k=1} s^2_{\nu_k}\right]^{-1}
\approx (4.4\ \mbox{TeV})^2.
\label{pubdi}
\eeq
In other words, expression (\ref{pub}) shows that
the unacceptable large-mass behaviour of the amplitudes ($\propto m^2_N$)
is actually cured when a sensible light-heavy mixing (at most 
$\propto m^{-2}_N$) is taken into account
\cite{delAguila:1982yu,Cheng:1991dy}. 

%=========================================================================
\begin{figure}[tbh]
\begin{center}
\epsfig{file=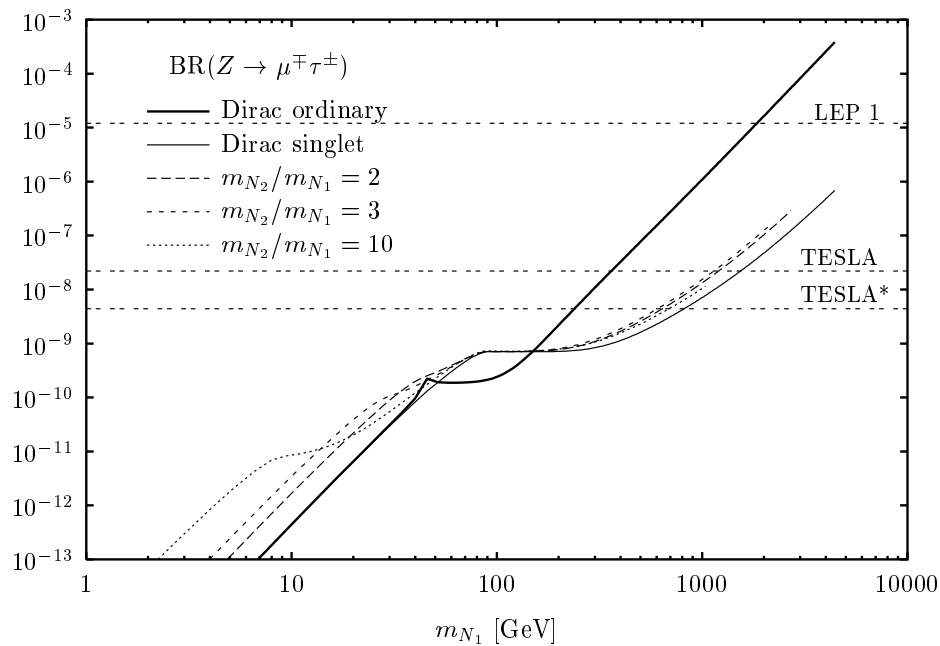,width=0.8\linewidth}
\end{center}
\caption{
\label{figure-bergmt}
{\it 
Upper limit of BR$(Z\to\mu^\mp\tau^\pm)$ assuming a light neutrino sector 
mixing with: 
(i) one heavy ordinary (thick-solid) or singlet (thin-solid) Dirac neutrino 
   of mass $m_{N_1}$; 
(ii) two heavy right-handed singlet Majorana neutrinos (dashed lines) with 
     masses $m_{N_1}$ and $m_{N_2}$.
}}
\end{figure}
%=========================================================================

For illustration, we show the less constrained results for 
$Z\to \mu^\mp\tau^\pm$ in Figure \ref{figure-bergmt} (thick-solid line), 
assuming the present (indirect) upper bounds on the corresponding mixings 
(\ref{lhmixnm}) and (\ref{lhmixnt}): given the heavy neutrino mass(es), the
branching fractions cannot exceed either the curves or the collider exclusion
limit. Regions below the curves correspond to mixings smaller than the 
upper bounds (\ref{lhmixne}) to (\ref{lhmixnt}).
Masses beyond the end points are acceptable
only if the mixings are smaller than the upper bounds.
Of course, since there are at least two unknowns, a neutrino mass and a 
combination of mixings, the LFV $Z$ decays cannot improve the 
bounds on the mixings without assuming a value for the heavy neutrino 
mass(es). This is in contrast to the LFV $\mu$ decays for sufficiently
heavy neutrinos (\ref{plateau}).

%=====================================================================
\section{\label{s-2majo}The LFV $Z$ decays in the $\nu$SM 
%with two heavy 
with right-handed Majorana singlets
}
%=====================================================================
Let us now consider the case when the heavy neutrinos are Majorana 
particles.
Actually this a very interesting possibility since such states 
belong to the particle content of most GUT theories, like SO(10).
Furthermore, they may participate in the seesaw mechanism,
that explains the smallness of the observed neutrino masses by
introducing a general Majorana neutrino \cite{Majorana:1937vz}
mass matrix, incorporating ordinary Dirac mass terms $m_D$, of a size 
typical to the charged lepton sector, and lepton-number violating Majorana 
mass terms at a higher scale $M_R \gg m_D$. Majorana mass terms 
$M_R \overline{\nu^c_R} \nu_R+\mbox{h.c.}$, {\em with $\nu_R$ being 
right-handed singlets} under the SM group, are gauge invariant, but violate 
lepton number by two units. The physical states after diagonalization of the 
mass matrix are, respectively, light and heavy Majorana neutrinos with masses
\bea
m_\nu\approx m^2_D/M_R,\quad m_N\approx M_R\gg m_\nu.
\eea 

If there is only {\em one} generation of heavy neutrinos, the light-heavy
mixings are fixed to be very small, 
\bea
s_\nu \approx m_D/M_R \approx \sqrt{m_\nu/m_N},
\eea
leading to unobservable LFV effects.

But this is not the case when one includes {\em several} right-handed Majorana 
neutrinos with inter-generation mixings 
\cite{Bernabeu:1993up,Korner:1993an,Ilakovac:1995kj}. 
We will focus on the most conservative case of {\em two heavy right-handed 
singlets}.

%===============================================
\subsection{\label{majo}LFV with Majorana neutrinos} 
%==================================================

Let us consider $n_G$ generations of charged leptons (Dirac fermions), whose 
left-handed components ($\ell^0_L = e_L,\mu_L,\tau_L,\dots$) belong to the 
same isodoublet as $n_G$ left-handed neutrinos 
($\nu^0_L = \nu_e,\nu_\mu,\nu_\tau,\dots$) and, in addition, $n_R$ 
right-handed neutrino singlets. The interaction eigenstates are a mixture 
of physical states given by 
\cite{Schechter:1980gr,Pilaftsis:1992ug,Ilakovac:1995kj} 
\bea
\label{lmix}
{\ell^0_L}_i &=& \dis\sum_{j=1}^{n_G} {\bf U}^{\ell_L}_{ij}\ {\ell_L}_j, 
\\
\label{numix}
{\nu^0_L}_i &=& \dis\sum_{j=1}^{n_G+n_R} {\bf U}_{ij}\ {\nu_L}_j,
\eea
where $\nu$=$\eta\ \nu^c$ are $n_G+n_R$ Majorana fields (i.e. self-conjugate 
up to a phase $\eta$). 

In the charged-current interactions, one  must replace the leptonic mixing 
matrix ${\bf V}$ by its generalized version, the rectangular 
$n_G\times(n_G\times n_R)$ matrix ${\bf B}$,
\bea
\label{def-b}
{\bf B}_{ij}\equiv\dis\sum^{n_G}_{k=1} {\bf U}^{\ell^*_L}_{ki}{\bf U}_{kj}.
\eea
Therefore, in the physical basis, 
\bea
\label{ccmix}
-{\cal L}_{CC}&=&\frac{g}{\sqrt{2}}W_\mu \overline{\ell^0_L}_i
\gamma^\mu P_L{\nu^0_L}_i + h.c. 
\nn\\
&=&
\frac{g}{\sqrt{2}}W_\mu\ {\bf B}_{ij}\ \overline{\ell_L}_i
\gamma^\mu P_L{\nu_L}_j + h.c.,
\eea
where $P_{R,L}=\frac{1}{2}(1\pm\gamma_5)$. 

But the main feature distinguishing Dirac and Majorana cases is the existence 
of {\em non-diagonal} $Z\nu_i\nu_j$ vertices (flavour-changing neutral current),
{\em coupling both left- and right-handed} components of the Majorana 
mass eigenstates to the $Z$ boson, 
\bea
\label{ncmix}
-{\cal L}^Z_{NC}
&=&
\frac{g}{2c_W}Z_\mu
[\overline{\nu^0_L}_i\gamma^\mu P_L {\nu^0_{L_i}} -
 \overline{\nu^{0c}_L}_i\gamma^\mu P_R {\nu^{0c}_{L_i}} ]
\nn\\
&=&
\frac{g}{2c_W}Z_\mu \overline{\nu_i} 
  [{\bf C}_{ij}\gamma^\mu P_L-{\bf C}^*_{ij}\gamma^\mu P_R] \nu_j,
\eea
where $\nu^{0c}_L=C\overline{\nu^0_L}^T$ is the charge-conjugate of $\nu^0_L$, 
which is right-handed, and
\bea
\label{def-c}
{\bf C}_{ij}\equiv\dis\sum^{n_G}_{k=1} {\bf U}^*_{ki}{\bf U}_{kj},
\quad (i,j=1,\dots,n_G+n_R),
\eea
a quadratic $(n_G+n_R)^2$ matrix.
Such flavour-changing NC vertices appear in graphs D1 and D3 of Figure 
\ref{fig-vertex} where Majorana neutrinos couple directly to the $Z$, and a $W$
or a Goldstone boson $\phi$ is exchanged:
\bea
v_W(i,j) & = & 
-{\bf C}_{ij}\big[ \lambda_Q(\c_{0}+\c_{11}+\c_{12}+\c_{23}) 
%\nn\\ &&\hspace{1cm}
-2\c_{24}+1\big]
%\nn \\&   & 
+ {\bf C}^*_{ij}\sqrt{\lambda_i\lambda_j}\ \c_0,
\\ 
v_{\phi}(i,j)  & = &
-{\bf C}_{ij}\displaystyle\frac{\lambda_i\lambda_j}{2}\c_0 
%\\   & & 
+ 
{\bf C}^*_{ij}\displaystyle\frac{\sqrt{\lambda_i\lambda_j}}{2}
 \left[\lambda_Q\c_{23}-2\c_{24}+\displaystyle\frac{1}{2}\right]. 
\eea
%The (diagonal) contributions of these graphs for Dirac virtual neutrinos
%[$v_W(i)$ and $v_\phi(i,j)$] are obtained by the replacements:
%\bea
%\c_{..}\equiv\c_{..}(\lambda_i,\lambda_j)&\to&\c_{..}(\lambda_i,\lambda_i),\\
%{\bf C}_{ij}&\to&  (v_i+a_i)\ \delta_{ij} = \delta_{ij},\quad \\
%{\bf C}^*_{ij}&\to& -(v_i-a_i)\ \delta_{ij} = 0.
%\eea 
The other diagrams remain unchanged compared to the Dirac case and the 
resulting form factor reads:\footnote{
We have compared our formulae with Eqn. (B1) of \cite{Pilaftsis:1995tf}
and found disagreement, in particular the appearance of a tensor integral
$\c_{22}$ at several instances. $\c_{22}$ is UV-finite and has no
numerical impact on the amplitudes for large neutrino masses, where
we find full agreement. However, the rearrangement of (\ref{VD})
leads to the well established Dirac vertices only when using our expressions.
%The same holds for \cite{Frank:1996xt} where the formulae of
%\cite{Pilaftsis:1995tf} are used, plus additional deviations.
}
\bea
\label{FMdef}
{\cal V}_{\rm Maj}(Q^2) &=&\sum^{n_G+n_R}_{i,j=1}
{\bf B}_{\ell_1 i} {\bf B}^*_{\ell_2 j}
 V_Z(i,j),
\\
V_Z(i,j)&=&v_W(i,j)+v_{\phi}(i,j)+v_{WW}(i)
%\nn\\&& 
+ v_{\phi\phi}(i)+v_{W\phi}(i)+v_{\Sigma}(i).
\label{FM}
\eea
We have used the Feynman rules in \cite{Denner:1992vz,Denner:1992me} to
properly handle interactions involving Majorana particles.
%They rely on the notion of fermion {\em flow} rather than that  of fermion {\em number flow}
%since the latter is not conserved for vertices with Majorana particles.
%Finally, for every such vertex one has to add also the reversed one
%while retaining the usual propagators.  
%Effectively, one gets the same result for the vertex $Wl_i\chi_j$ but a 
%{\em factor two} larger for the vertex $Z\chi_i\chi_j$ in comparison to the 
%case of Dirac neutrinos.

It turns out convenient to cast (\ref{FM}) as
\bea
V_Z(\lambda_i,\lambda_j)&=&\delta_{ij}F(\lambda_i)+{\bf
C}_{ij}G(\lambda_i,\lambda_j)
%\nn\\
+ {\bf C}^*_{ij} \sqrt{\lambda_i\lambda_j} H(\lambda_i,\lambda_j).
\label{VM}
\eea
The Dirac vertex function (\ref{l24}) is then 
\bea
V_Z(\lambda_i)=F(\lambda_i)+G(\lambda_i,\lambda_i).
\label{VD}
\eea 
The form factor (\ref{FMdef}) is UV-finite, but the vertex function 
$V_Z(\lambda_i,\lambda_j)$ is not.  The divergences are such 
that they exactly cancel due to unitarity relations among the mixing 
matrix elements of ${\bf B}$ and ${\bf C}$ 
\cite{Ilakovac:1995kj,Illana:1999ww}.
The same relations allow to write  ${\cal V}_{\rm Maj}$  in terms of only the 
heavy sector, assuming the light sector  being massless:
\bea
\label{larandsm}
 {\cal V}_{\rm Maj}(Q^2) 
 &=&\sum^{n_R}_{i,j=1}{\bf B}_{\ell_1 N_i} {\bf B}^*_{\ell_2 N_j}
\nn\\
&&\times ~\Big\{\delta_{N_i N_j}
[\ F(\lambda_{N_i})-F(0)+G(\lambda_{N_i},0)
%\nonumber\\ &&\quad\quad\quad\quad 
+ G(0,\lambda_{N_i}) - 2 G(0,0)]  
\nonumber\\
&&+~{\bf C}_{N_i N_j}
[\ G(\lambda_{N_i},\lambda_{N_j})-G(\lambda_{N_i},0)
%\nonumber\\&&\quad\quad\quad\quad
- G(0,\lambda_{N_j})+G(0,0)] 
\nonumber\\
&&+~{\bf C}^*_{N_i N_j} \sqrt{\lambda_{N_i}\lambda_{N_j}} \
H(\lambda_{N_i},\lambda_{N_j})\Big\}.
\eea

%===========================================================
\subsection{\label{smplt}The $\nu$SM with two heavy Majorana singlets
%Majorana neutrinos ($n_R=2$)
}
%===========================================================
In the simple case of $n_R=2$ heavy right-handed singlet neutrinos $N_1$ and
$N_2$, mixing with a massless sector, the ${\bf B}$ and ${\bf C}$ matrices are 
fully determined by the ratio of the two physical heavy masses squared
$r\equiv m^2_{N_2}/m^2_{N_1}$ and the light-heavy mixings $s^2_{\nu_\ell}$,  
here 
\beq
s^2_{\nu_\ell} \equiv \sum_{i} |{\bf B}_{\ell N_i}|^2.
\label{lihe}
\eeq
Their explicit values are \cite{Ilakovac:1995kj}:
\bea
{\bf B}_{\ell N_1} &=& \frac{r^{1/4}}{\sqrt{1+r^{1/2}}} s_{\nu_\ell} ,
\\
{\bf B}_{\ell N_2} &=& \frac{i}{\sqrt{1+r^{1/2}}} s_{\nu_\ell} ,
\\
{\bf C}_{N_1 N_1} &=& \frac{r^{1/2}}{1+r^{1/2}}\sum_{\ell} s^2_{\nu_\ell},
\\
{\bf C}_{N_2 N_2} &=& \frac{1}{1+r^{1/2}}\sum_{\ell} s^2_{\nu_\ell},
\\
{\bf C}_{N_1 N_2}\ =\ -~{\bf C}_{N_2 N_1} &=& \frac{ir^{1/4}}{1+r^{1/2}}
                                        \sum_{\ell} s^2_{\nu_\ell}.
\eea
The mass ratio $r$ is a free parameter and the light-heavy mixings
are constrained by present experiments as shown in Section \ref{s-dirac}.
Upper values for the branching ratios of $Z\to\mu^\mp\tau^\pm$,  
obtained from the experimental bounds given the heavy masses $m_{N_1}$, 
$m_{N_2}$, are also displayed in Figure \ref{figure-bergmt}.

The case of two equal-mass Majorana neutrinos $m_{N_1}=m_{N_2}$ is equivalent 
to one heavy singlet Dirac neutrino,\footnote{In fact, two equal-mass Majorana 
neutrinos with opposite CP parities form a Dirac neutrino.} and it approaches 
rapidly the ordinary Dirac case for small masses. This phenomenon is just 
another example of the ``practical Dirac-Majorana confusion theorem''  
\cite{Kayser:1982br} (see also the recent discussion in
\cite{Zralek:1997sa,Czakon:1999cd} and references therein). 
If both masses $m_{N_1}$ and $m_{N_2}$ are small, the amplitude goes as
$\sqrt{r}\lambda_{N_1}=m_{N_1} m_{N_2}/M^2_W$ with the same global
factor $a_1$ as in the ordinary Dirac case (\ref{a1reim}). This can been
seen in Figure \ref{figure-bergmt} not far below the $Z$ peak, where the
branching ratios grow with $\lambda^2_{N_1}$ and scale with the ratio of
the two neutrino masses squared.

If one of the neutrinos has the mass of the $Z$ boson, the imaginary parts of 
the amplitudes ${\cal V}_{\rm Dir,Maj}$ (coming from the
subtraction(s) at $\lambda_N=0$) dominates, both for the Dirac and the
Majorana cases.   
This happens since the real parts are slowly varying for $M_N \leq M_Z$, while the
imaginary parts vanish for $M_Z < M_N + M_{N'}$.
Further, since these 
imaginary parts necessarily come from accounting the subtractions of the zero 
mass limits, they are independent of the value of $r$. This results in
common values of the branching ratios for $m_{N_1}=M_Z$ for any value of
$m_{N_2}$. Nevertheless, the subtraction of the light sector implied by the 
unitarity constraints is not the same for the cases of a heavy ordinary Dirac 
neutrino and heavy Majorana singlets. One finds explicitly
\bea
\frac{\Im{\rm m} ({\cal V}_{\rm Dir})}{s_{\nu_{\ell_1}}s_{\nu_{\ell_2}}} 
= -1.0524, 
\quad
\frac{\Im{\rm m} ({\cal V}_{\rm Maj})}{s_{\nu_{\ell_1}}s_{\nu_{\ell_2}}} 
= -2.0653.
\eea

The expansion of the form factor (\ref{larandsm}) in the large neutrino mass 
limit $\lambda_{N_1}\gg 1$, at fixed $r$, leads to (see Appendix 
\ref{appsub-maj}):
\bea
{\cal V}_{\rm Maj}(Q^2)
&=&s_{\nu_{\ell_1}} s_{\nu_{\ell_2}} 
%\hspace{2.8cm}
%\nn\\ \times
\Bigg\{\frac{\sum_\ell s^2_{\nu_{\ell}}}{(1+r^\frac{1}{2})^2}
\left(\frac{3}{2}r+\frac{r^2+r-4r^\frac{3}{2}}{4(1-r)}
\ln r\right)\lambda_{N_1} 
\nn\\
&&+~\frac{1}{2}\left(3-\frac{1-2s^2_W}{6}\lambda_Q\right)
\ln\lambda_{N_1}\Bigg\} + {\cal O}(1),
\label{majlim}
\eea
that agrees with \cite{Ilakovac:1995kj} for the unphysical value $\lambda_Q=0$.
The constant in front of the leading term coincides for $r=1$ with the ordinary
Dirac case, except for an extra damping factor $\sum_\ell s^2_{\nu_\ell}$, that
makes the Dirac singlet case in particular, and the Majorana case in general, 
more sensitive to the present bounds on the light-heavy mixings.
The constant in front of the $\ln\lambda$ term, subleading but not so much 
mixing-suppressed, is identical to the one in the ordinary Dirac case 
(\ref{vbig}). 

We have cut again the curves at the perturbative unitarity mass limits.
Due to the different number of degrees of freedom of the Majorana
particles, the condition on the heavy neutrino width is this time 
\cite{Fajfer:1998px},
\bea
\Gamma_{N_i}\simeq 2\times
\frac{\alpha_W}{8M^2_W}m^3_{N_i}\sum_{\ell}|{\bf B}_{\ell N_i}|^2\leq 
\frac{1}{2} m_{N_i},
\label{kk}
\eea
resulting in the mass limits
\beq
\label{majolim}
m^2_{N_1}\equiv\frac{1}{r} m^2_{N_2} 
&\lsim& \frac{1+r^{1/2}}{r}\times (3.1\ \mbox{TeV})^2.
\eeq

We see from the figure that GigaZ has a discovery
potential, preferentially in the large neutrino mass
region, if the light heavy-mixings are not much below the present
upper limits. Due to the different coupling structure, the simple 
sequential Dirac neutrino case does not constitute a limiting case for 
large masses.

%=====================================================================
\section{\label{s-concl}Concluding remarks   
}
%=====================================================================
The sensitivity of the GigaZ mode of the future TESLA linear collider
to rare, lepton-flavour violating $Z$ decays has been studied. 
We have determined the full one-loop expectations for the {\em direct} 
%study of the 
lepton-flavour changing
process $Z\to\bar\ell_1\ell_2$ with virtual Dirac or Majorana
neutrinos.
This is an interesting theoretical issue in view of the evidences for
tiny neutrino masses from astrophysics, which might be also indicative for
the existence of heavy neutrinos in some Grand Unifying Theory.  
Both the exact analytical form and the large and
small neutrino mass limits of the branching ratios are given, thereby
cross-checking the existing literature. 
%From our study of the LFV decays of the $Z$ boson 
From our numerical studies, taking into account the present
experimental results, we conclude that:
(i) the contributions from the observed light neutrino sector are
far from experimental verification (BR$\lsim 10^{-54}$);
(ii)  the GigaZ mode of the future TESLA linear
collider, sensitive down to  about BR $\sim 10^{-8}$, 
might well have a chance to produce such processes, if heavy neutrinos
exist in Nature and if they mix with the light ones in a sizeable way.
Finally, we have shown that we could gain from observation of the LFV $Z$ decays 
alternative informations compared to the LFV $\mu$ decays.

%  \section{Not quoted so far}
%  
%  \cite{%
%  Frank:1998ih},
%  \cite{%
%  Couture:1997qq},
%
%  \cite{%
%  Doncheski:1989ai}

\section*{Acknowledgements}

We would like to thank 
F.~del Aguila,
J.~Gluza,
A.~Hoefer,
A.~Pilaftsis,
G.~Wilson
 for helpful discussions and 
M.~Jack for a fruitful cooperation in an earlier stage of the project. 
This work has been supported in part by CICYT,
by Junta de Andaluc{\'\i}a under contract FQM 101, 
and by the European Union HPRN-CT-2000-00149.

\appendix
%=====================================================================
\section{
\label{a-vert} 
Tensor integrals and vertex functions
}
%=====================================================================
We have introduced dimensionless two- and three-point one-loop 
functions:
\begin{eqnarray}
\label{b1-def}
{\b}_1(\lambda_i) &\equiv& B_1(0;m^2_i,M^2_W),
\\
\label{cbar-def}
\bar{\c}_{..}(\lambda_i) 
%\nn\\&=&
&\equiv& M^2_W\ C_{..}(0,Q^2,0;m^2_i,M^2_W,M^2_W),
\\
\label{c-def}
\c_{..}(\lambda_i,\lambda_j) 
%\nn\\ &=&
&\equiv& M^2_W\ C_{..}(0,Q^2,0;M^2_W,m^2_i,m^2_j),
\end{eqnarray}
from the usual loop integrals \cite{'tHooft:1972fi,Passarino:1979jh}
with the tensor decomposition (Minkowski metric):
\ba
\label{B-def}
B^\mu(p^2;m^2_0,m^2_1)&=&p^\mu B_1, 
\\
\label{cmu-def}
C^\mu(p^2_1,Q^2,p^2_2;m^2_0,m^2_1,m^2_2)&=&p^\mu_1 C_{11} + p^\mu_2 C_{12},
\\
\label{cmunu-def}
C^{\mu\nu}(p^2_1,Q^2,p^2_2;m^2_0,m^2_1,m^2_2)
&=&p^\mu_1 p^\nu_1 C_{21} + p^\mu_2 p^\nu_2 C_{22} 
+ (p^\mu_1 p^\nu_2 + p^\mu_2 p^\nu_1) C_{23}
%\nn\\ &&
+ g^{\mu\nu} M^2_W C_{24}. \nn\\
\ea
The tensor integrals are numerically evaluated  with the computer
program {\tt LoopTools} \cite{Hahn:1998yk}, based on {\tt FF}
\cite{vanOldenborgh:1990yc}.
All the numerical results for the Dirac case have been carefully
checked against an older approach described in \cite{Mann:1984dvt}. 

The following definitions of the integrals in $d$ dimensions are 
useful:
\ba
\label{uv-b1}
\b_1(\lambda_i)&=& -\frac{\Delta_\epsilon}{2} 
+ \int_0^1  dx \ x \ln[(1-\lambda_i)x+\lambda_i-i\epsilon] ,
\\
\label{uv-c24}
\c_{24}(\lambda_i,\lambda_j) &=&\frac{\Delta_\epsilon}{4}
-\frac{1}{2}\int_0^1 dx 
\int_0^x dy ~\ln  D_{ijW},
\\
\label{co123}
\c_{0,11,23}(\lambda_i,\lambda_j) &=& - \int_0^1 dx \int_0^x dy 
~[1,-y,y(1-x)]~\frac{1}{D_{ijW}},
\ea
with $\lambda_i=m^2_i/M^2_W$, $\Delta_\epsilon=2/(4-d)-\gamma+\ln4\pi$ and
\ba
 \label{dijw}
%D(\lambda_i,\lambda_j,,1,\lambda_Z) = 
D_{ijW} \equiv&
\lambda_Z xy + (1-\lambda_j)x+ [-\lambda_Z+(\lambda_i-1)]y + \lambda_j -
i\epsilon.
\ea
To get the barred tensor integrals $\cbar$, replace $D_{ijW}$ by:
\ba
\label{dwwi}
%{{\bar D}}  = 
D_{WWi} \equiv  
\lambda_Z xy - (1-\lambda_i)x+ [-\lambda_Z-(\lambda_i-1)]y +1 -
i\epsilon.
\ea

The functions $\b_1$, $\c_{24}$ and $\cbar_{24}$ are UV-divergent but
the physical amplitudes are finite.

%=======================================================================
\subsection{\label{a-zeromass}Light neutrino mass expansions}
%=======================================================================
%Some more details are given in Appendix D.2 of \cite{Illana:1999ww}.
Let us first list the value of the necessary tensor integrals for
massless neutrinos and $\lambda_Q\ne 0$ \cite{Mann:1984dvt}:
\ba
\b_1(0)&=&-\frac{\Delta_\epsilon}{2}-\frac{1}{4},\\
\c_0(0,0)&=&-c_0,\\
\c_{11}(0,0)&=&-\frac{1}{\lambda_Q}(c_0-1+\ln\lambda_Q-i\pi),\\
\c_{12}(0,0)&=&\c_{11}(0,0),\\
\c_{23}(0,0)&=&-\frac{1}{\lambda_Q^2}\left[(\lambda_Q+2)c_0-\frac{\lambda_Q}{2}
               -2+2(\ln\lambda_Q-i\pi)\right],\\	       
\c_{24}(0,0)&=&\frac{\Delta_\epsilon}{4}+\frac{1}{4\lambda_Q}[-2(\lambda_Q+1)c_0
	       +3\lambda_Q+2-(\lambda_Q+2)(\ln\lambda_Q-i\pi)],\hspace{4mm}\\
\cbar_0(0)&=&-\bar c_0,\\
\cbar_{11}(0)&=&\frac{1}{\lambda_Q}(\bar c_0-B+1),\\
\cbar_{12}(0)&=&\cbar_{11}(0),\\
\cbar_{23}(0)&=&-\frac{2}{\lambda_Q^2}
               \left(\bar c_0-B+1-\frac{\lambda_Q}{4}\right),\\
\cbar_{24}(0)&=&\frac{\Delta_\epsilon}{4}-\frac{1}{2\lambda_Q}\left[
                \bar c_0-B+1-\frac{3}{2}\lambda_Q+\pi\lambda_Qy-2\lambda_Qy
		\arctan(2y)\right],
\ea	     
with
\ba
\label{core}
\lambda_Q~c_0 &=& \frac{\pi^2}{6} - \litwo\left(\frac{1}{1+\lambda_Q}\right)
-\frac{1}{2}\ln^2(1+\lambda_Q) + \pi \ln(1+\lambda_Q) \  i,
\\
\label{barcore}
\lambda_Q~{\bar c}_0 &=& \frac{\pi^2}{6} - \litwo(1-\lambda_Q) 
% next line a misprint in Annalen d. Physik. 19-11-99
+2 \Re e \litwo\left[(\lambda_Q-1)\left(\frac{\lambda_Q}{2}-1+\lambda_Q y \ i\right)
\right]
\nl &&
-~   2 \Re e \litwo\left(1-\frac{\lambda_Q}{2}-\lambda_Q y \ i\right),
\\
\label{B}
B &=& 2y\left[\arctan(2y) + \arctan\left(\frac{\lambda_Q-1}{3-\lambda_Q}2y 
\right)\right]. 
\ea	     
%valid in the range $0<\lambda_Q<4$.

After expanding the tensor integrals for small neutrino masses
(see Appendix D.2 of \cite{Illana:1999ww}), the vertex function for the
case of a light Dirac neutrino reads:
\ba
\label{small}
V_Z(\lambda \ll 1; \lambda_Q\ne0) &=&
%I_3^{\nu_L} \frac{8c_W^2}{D-4}
%+
%a_0  
V_Z(0; \lambda_Q)
+ a_1 \lambda + {\cal O}(\lambda^2),
\ea
where the terms proportional to $\lambda\ln\lambda$ {\em have cancelled out} 
and
% \cite{Mann:1982xw}
\ba
\label{small-vdi}
V_Z(0;\lambda_Q) &=&
%I_3^{\nu_L} 
%\frac{8c_W^2}{D-4}
-2c^2_W\Delta_\epsilon
%+ V'(\lambda_i \ll 1, \lambda_Z),
%\\
%V'(\lambda_i \ll 1, \lambda_Z) &=& 
%%%%% general expression for $a_0$ is (unpublished so far, see notes
%   N82p.71,16-2-82-3 and corrected p.73):
+ \frac{2+3\lambda_Q}{2\lambda_Q}(\ln
\lambda_Q-\pi \ i) - \frac{1}{4\lambda_Q^2}(7\lambda_Q^2+14\lambda_Q-8)
\frac{(1+\lambda_Q)^2}{\lambda_Q}c_0 
 \nl && +~\frac{2}{\lambda_Q^2}(1+2\lambda_Q)({\bar c}_0-B)
% term has wrong factor on page 71: +\frac{2}{3}[\pi y - 2y \arctan(2y)] 
+\frac{6}{\lambda_Q}[\pi y - 2y \arctan(2y)] .\hspace{8mm}
% \nl && 
\ea
Only the functions ${\cbar}_{..}$ develop imaginary parts, and only for 
$\lambda_Q > 4\lambda_i$. At the $Z$ peak the numerical result is:
\ba
\label{a0}
V_Z(0; \lambda_Z) &=&
%I_3^{\nu_L} \frac{8c_W^2}{D-4}
-2c^2_W\Delta_\epsilon
+
%a_0  
1.2584 + 1.0524\ i.
\ea
The linear term in the expansion (\ref{small}) has the coefficient
\cite{Riemann:1982rq,Illana:1999ww,Illana:2000zy}:
\ba
\label{a1ana}
a_1(\dz) &=&
-\frac{2}{\dz}(1+\dz) c_0 
+ \frac{1}{2\dz^2}(4\dz^2-5\dz-6){\bar c}_0-\frac{2}{\dz}
( \ln \dz -\pi \ i)
\nl &&
+~\frac{1}{8\dz^2} (25\dz^2-38\dz-24) +\frac{1}{2\dz}(2-\dz)\pi y  
\nl && +~
\frac{1}{\dz^2}(\dz^2+7\dz+6)y\arctan (2y)  
\nl &&+~ \frac{3}{\dz^2} (3\dz+2) 
y\arctan\left(\frac{\dz-1}{3-\dz}2y\right)\\
&=&2.5623  - 2.2950  \ i.
\ea
%with $y$ from (\ref{y}).
%Numerically, (\ref{a1reim}) results for  $\lambda_Z = 1/c^2_W$.
% \ba
% \label{a1}
 %       a_1 &=&  2.5623  - 2.2950  \ i.
% \ea

The behaviour of (\ref{small}) is {\em in contrast to the case} 
$\lambda_Q=0$ for which \cite{Ma:1980px,Mann:1984dvt,Illana:1999ww}:
\ba
V_Z(\lambda\ll 1;\lambda_Q=0)
= \frac{1}{2}( -4c^2_W\Delta_\epsilon + 6\lambda + 2\lambda\ln\lambda )
 + {\cal O}(\lambda^2).
\ea

%=====================================================================
\subsection{\label{appsub-maj}Heavy neutrino mass expansions}
%=====================================================================
The limits of the necessary tensor integrals and the vertex function 
in the Dirac case for large neutrino masses can be found in 
Appendix D of \cite{Illana:1999ww}. We collect below the large
mass expansions of the tensor integrals that are also needed for the Majorana
case, namely one or two identical neutrinos running in the loop:
\ba
\b_1(\lambda_i) &=&
-\frac{\Delta_\epsilon}{2}
+ \frac{1}{2}\ln\lambda_i 
-\frac{3}{4} + \frac{\ln\lambda_i}{\lambda_i} - \frac{1}{2\lambda_i}
+{\cal O}(1/\lambda_i^2),
\\
\cbar_{0}(\lambda_i) &=& -\frac{\ln\lambda_i}{\lambda_i} -
\left[1-4y\arctan\left(\frac{1}{2y}\right) \right]
\frac{1}{\lambda_i} +{\cal O}(1/\lambda_i^3),
\\
\cbar_{11}(\lambda_i)=\cbar_{12}(\lambda_i)
&=&\frac{1}{2}\frac{\ln\lambda_i}{\lambda_i}
+{\cal O}(1/\lambda_i^2),\\
\cbar_{23}(\lambda_i)&=&-\frac{1}{6}\frac{\ln\lambda_i}{\lambda_i}
+{\cal O}(1/\lambda_i^2),
\\
\cbar_{24}(\lambda_i) &=& 
\frac{\Delta_\epsilon}{4}
-\frac{1}{4}\ln\lambda_i
+\frac{3}{8}
+ \left(-6+\lambda_Q\right)\frac{\ln\lambda_i}{12\lambda_i}
\nl&&+\left[-30+5\lambda_Q+24(4-\lambda_Q)y\arctan\left(\frac{1}{2y}\right) \right]
\frac{1}{72\lambda_i}\nonumber \\ &&
+{\cal O}(1/\lambda_i^2),
\\
\c_0(\lambda_i,\lambda_i) &=& -\frac{1}{\lambda_i} 
+\frac{\ln\lambda_i}{\lambda_i^2}
-\left(12+\lambda_Q\right)\frac{1}{12\lambda_i^2} +{\cal O}(1/\lambda_i^3),
\\
\c_{11}(\lambda_i,\lambda_i)=\c_{12}(\lambda_i,\lambda_i)
&=&\frac{1}{4\lambda_i}+{\cal O}(1/\lambda_i^2),
\\
\c_{23}(\lambda_i,\lambda_i) &=&-\frac{1}{18\lambda_i}+{\cal O}(1/\lambda_i^2),
\\
\c_{24}(\lambda_i,\lambda_i) &=&
\frac{\Delta_\epsilon}{4}
- \frac{1}{4}\ln\lambda_i 
+\frac{1}{8} 
+ \left(-9+\lambda_Q\right)\frac{1}{36\lambda_i}+{\cal O}(1/\lambda_i^2),
\hspace{8mm}
\ea
with
\ba
y = \sqrt{1/\lambda_Q - 1/4}.
\ea
Substituting the expressions above in (\ref{l24}) one gets the Dirac vertex 
function of (\ref{vbig}).

Besides, we need some additional expansions for two Majorana fermions with 
different large masses masses $\lambda_i\ne\lambda_j$,
\ba
% 17 02 2000 p.3
\c_0(\lambda_i,\lambda_j) &=&
- \frac{1}{\lambda_i - \lambda_j}\left[
\frac{\lambda_i}{\lambda_i -1}\ln\lambda_i 
- 
\frac{\lambda_j}{\lambda_j -1}\ln\lambda_j
\right]
\nl
% the sign was wrong of next terms 07-03-00 due to sign error in form
% file c0ijw.frm 
% && -~\frac{\lambda_Q}{(\lambda_i - \lambda_j)^2} 
  && +~\frac{\lambda_Q}{(\lambda_i - \lambda_j)^2} 
\left[
1
+\frac{1}{2}\left(1-\frac{2\lambda_i}{\lambda_i -
\lambda_j}\right)\ln\lambda_i 
+\frac{1}{2}\left(1+\frac{2\lambda_j}{\lambda_i -
\lambda_j}\right)\ln\lambda_j
\right]
\nl && +~ {\cal O}(1/\lambda^3),
% derived in:
% delta_z terms in c0ijw.frm, where also the limit i=j is checked.
\ea
and, to a lower accuracy in the expansion parameters:
\ba
\c_{11}(\lambda_i,\lambda_j) &=& \frac{1}{2} \frac{1+\ln \lambda_i}{\lambda_i - \lambda_j}
+
\frac{\lambda_j \ln \lambda_j-\lambda_i\ln \lambda_i}{2(\lambda_i - \lambda_j)^2}
+ {\cal O}(1/\lambda^2)
,
\\
\c_{12}(\lambda_i,\lambda_j) &=&\c_{11}(\lambda_j,\lambda_i),
\\
\c_{23}(\lambda_i,\lambda_j) &=& - \frac{1}{6(\lambda_i - \lambda_j)^2} 
\left[ (\lambda_i + \lambda_j) 
% corr. by JoIl 05/00:
%- \frac{\lambda_i \lambda_j}{\lambda_i -
 - \frac{2\lambda_i \lambda_j}{\lambda_i -
\lambda_j} \left(\ln \lambda_i - \ln \lambda_j \right) \right] 
%\nl && +~ 
+{\cal O}(1/\lambda^2),
%\ea
%Finally:
%\ba
% 17-02-00-c
\\
\label{c24limij}
\c_{24}(\lambda_i,\lambda_j) &=& 
\frac{3}{8} - \frac{1}{4} \frac{(\lambda_i+1)\ln \lambda_i -
(\lambda_j+1)\ln \lambda_j}{\lambda_i - \lambda_j} 
%\nl &&-~
-\frac{\lambda_Q}{2} \c_{23}(\lambda_i,\lambda_j) + {\cal O}(1/\lambda^2). 
\hspace{12mm}
\ea
Actually,  $\c_{11}$ and $\c_{12}$ are irrelevant for large neutrino 
masses.

Finally, in (\ref{larandsm}) we need loop integrals where one neutrino mass is
large and the other one vanishes. They are all irrelevant except $\c_{24}$ 
in this limit, but we show their expansions for completeness: 
% Jose p. 27
\ba
\label{one1}
\c_0(\lambda,0) 
%= C_0(0,\lambda)  
&=&  - \frac{\ln \lambda}{\lambda} -
\frac{\ln \lambda}{\lambda^2} 
%\nl &&
+  \frac{\lambda_Q}{2\lambda^2}
+ {\cal O}(1/\lambda^3),
\\
\label{one2}
\c_{12}(0,\lambda)  =
\c_{11}(\lambda,0) 
&=&  \frac{1}{2\lambda} 
+(3+\lambda_Q)\frac{1}{6\lambda^2} 
 -\frac{\ln \lambda}{\lambda^2} + {\cal O}(1/\lambda^3),
\\
\label{one3}
\c_{11}(0,\lambda)  =
\c_{12}(\lambda,0) 
&=&  \frac{1}{2\lambda} 
+ (3+2\lambda_Q)\frac{\ln \lambda}{6\lambda^2} + \frac{
\lambda_Q}{3\lambda^2}+ {\cal O}(1/\lambda^3), 
\\
\label{one4}
\c_{23}(\lambda,0) 
%= C_{23}(0,\lambda)  
&=& -\frac{1}{6\lambda} 
-(2+\lambda_Q)\frac{1}{12\lambda^2} -\frac{\ln \lambda}{6\lambda^2}
+ {\cal O}(1/\lambda^3),
\\
\label{one5}
\c_{24}(\lambda,0) 
%= C_{24}(0,\lambda)  
&=& \frac{\Delta_{\epsilon}}{2} 
+ \frac{3}{8} - \frac{\ln \lambda}{4} 
%\nl &&
- \frac{\lambda_Q}{12\lambda}+
{\cal O}(1/\lambda^2),
\ea
and
\ba
\label{one0}
\c_{0,23,24}(0,\lambda) = \c_{0,23,24}(\lambda,0).
\label{one0p}
\ea

%For completeness, we give also the $\c$ functions
%for
%the case of three arbitrary  virtual particles.
%Then, the introduction of barred functions is not needed, and
%$D_{ijW}$ gets replaced by:  
%\ba
%\label{susy1}
%D_{ij3} &=& 
%\lambda_Z xy + (\lambda_3-\lambda_j)x+ [-\lambda_Z+(\lambda_i-\lambda_3)]y + \lambda_j -
%i\epsilon
%\nl
%&=&
%\lambda_3  ~  D(L_i,L_j,1,L_Z),
%\ea
%with
%\ba
%\label{susy2}
%L_A &=& \frac{\lambda_A}{\lambda_3}, ~~~ A=i,j,Z . 
%\ea
%Then it follows:
%\ba
%\label{susy5}
%\c_{0,11,23} 
%&=& 
%- \frac{1}{\lambda_3} \int_0^1 dx \int_0^x dy \frac{[1,-y,y(1-x)]}{D(L_i,L_j,1,L_Z)},
%\\
%\label{susy6}
%\c_{24}&=& - \frac{1}{2(D-4)} - \frac{1}{2} \int_0^1 dx \int_0^x dy 
%\left[  \ln \lambda_3 + \ln D (L_i,L_j,1,L_Z,) \right].
%\ea

%\begin{thebibliography}
\providecommand{\href}[2]{#2}
%\end{thebibliography}

%=======================================================================

\end{document}